\documentclass[12pt,preprint]{aastex}

\shorttitle{Constraints on Hadronic Models for GRB 090510}
\shortauthors{Asano, Guiriec \& M\'esz\'aros}

\begin{document}

\title{
Hadronic Models for the Extra Spectral Component in the short GRB 090510}
\author{\scshape Katsuaki Asano\altaffilmark{1},
Sylvain Guiriec\altaffilmark{2}, and
Peter M\'esz\'aros\altaffilmark{3}}
\email{asano@phys.titech.ac.jp, sylvain.guiriec@nasa.gov, nnp@astro.psu.edu}

\altaffiltext{1}{Interactive Research Center of Science, 
Tokyo Institute of Technology, 2-12-1 Ookayama, Meguro-ku, Tokyo 152-8550, Japan}
\altaffiltext{2}{University of Alabama in Huntsville, Huntsville, AL 35899}
\altaffiltext{3}{Department of Astronomy \& Astrophysics;
Department of Physics;
Center for Particle Astrophysics;
Pennsylvania State University,
University Park, PA 16802}

\date{Submitted; accepted}

\begin{abstract}
A short gamma-ray burst GRB 090510 detected by {\it Fermi} shows an extra 
spectral component between 10 MeV and 30 GeV, an addition to a more usual
low-energy ($<10$ MeV) Band component.
In general, such an extra component could originate from accelerated protons.
In particular,
inverse Compton emission from secondary electron-positron pairs and 
proton synchrotron emission are competitive models for reproducing
the hard spectrum of the extra component in GRB 090510.
Here, using Monte Carlo simulations, we test the hadronic scenarios against
the observed properties.
To reproduce the extra component around GeV with these models,
the proton injection isotropic-equivalent luminosity is required
to be larger than $10^{55}$ erg$/$s.
Such large proton luminosities are a challenge for the hadronic models.
\end{abstract}


\maketitle

\section{Introduction}
\label{sec:intro}
The very high bulk Lorentz factor and high magnetic field strengths
in outflows of gamma-ray burst (GRB) make GRBs potential sources of
ultra-high-energy cosmic rays \citep[UHECRs;][]{wax95,vie95,gialis2004}.
Recent time-resolved detections of GeV photons with {\it Fermi}/LAT
from GRB 080916C and GRB 090510 \citep{abd09,abd09b}
reveal that the minimum Lorentz factor $\Gamma_{\rm min}$, required to make
the sources optically thin for such GeV photons, should be
$\gtrsim 1000$. While those high values of $\Gamma_{\rm min}$ are favorable
to accelerate protons to very high energies, it makes difficult to produce
the photomeson-induced secondary photons and neutrinos frequently discussed by
many authors \citep[e.g.,][]{rac98,der02,asa05,der06,asa06,gup07}.
This is because, given the variability timescale $\Delta t$,
photon luminosity, and spectrum,
a higher bulk Lorentz factor $\Gamma$ decreases
the photon number density as $n_\gamma \propto \Gamma^{-5}$.
The wide-band spectra (10 keV - GeV) of GRB 080916C are well fitted by
a smooth Band function \citep{ban93}, and in that case the lack of
evidence for an extra spectral component is consistent with a low efficiency of
photomeson production due to high $\Gamma$.

However, from 0.5 s to 1.0 s after the
trigger time ($T_0$), the short GRB 090510 does exhibit
a very significant ($\geq 5\sigma$) spectral deviation to the standard Band function.
With the parameters of the standard Band function at
$\varepsilon_{\rm peak}=3.94 \pm 0.28$ MeV,
$\alpha=-0.58 \pm 0.06$, and $\beta=-2.83 \pm 0.20$,
the excess is adequately fit with an additional power law of index $-1.6$.
The power law extends from the lowest energy in
{\it Fermi}/GBM (i.e., 8 keV), to at least 31 GeV in {\it Fermi}/LAT.
The extra-component counts for about 37\% of the total fluence
and the powerlaw dominates the standard integrated-Band spectrum 
up to a few tens of keV and above 10 MeV.
This additional component is the most intense from $T_0+0.6$ s to $T_0+0.8$ s.
There is no clear evidence of existence of an extra component at times after $T_0+0.8$ s,
however there is not enough statistical significance in the data to fully reject this
hypothesis.
The onset of the main emission above 100 MeV is delayed about 200 ms compared to
the main spike from 8 keV up to few MeV. No lags have been found below 1 MeV
with respect to the lowest energy band 8-40keV over the whole light curve
(~$T_0$ to $T_0+1.5$ s). Above 1 MeV, lags increase progressively to reach
~248 ms and remaining constant after ~40 MeV.
This extra component could be of leptonic origin, e.g.,
external Compton emission from internal shocks
\citep[e.g.,][]{tom09} or synchrotron self-Compton emission from
the reverse shock or forward shock in the early afterglow phase \citep{ghi09} as
discussed for GRB 941017 \citep{gra03,pee04}.

In this Letter we consider an alternative possibility,
namely hadronic models, in particular
photomeson cascade and proton synchrotron models \citep{vie97,tot98,raz09}, for
representing the extra spectral component of GRB 090510,
which is present only in the prompt emission between
$T_0$ and $T_0+0.8$ s.
Although the cascade processes initiated by $p + \gamma \to p$/$n + \pi^0$/$\pi^+$
are complicated, the resultant photon signatures of proton cascades
mostly appear as synchrotron or inverse Compton emission
from secondary electron-positron pairs produced via $\gamma + \gamma \to e^+ + e^-$.
In such pair cascade processes the effective injection index of secondary pairs
tends to be about $-2$ \citep[e,g,][]{cop92} so that the synchrotron radiation
from secondary pairs yields a flat spectrum in an $\varepsilon^2 N(\varepsilon)$ plot,
while the power law index of the extra component in GRB 090510 is $\sim -1.6$.
However, the inverse Compton component from secondary pairs
can harden the spectrum as \citet{asa09} showed.
Using the numerical code of Monte Carlo techniques
in \citet{asa09} \citep[see also][]{asa07}, we constrain
the hadronic models in this Letter.

After a short discussion of our methods in \S \ref{sec:model},
we discuss the possibility of photon emission due to accelerated protons
for GRB 090510 in \S \ref{sec:results}, and summarize the results
in \S \ref{sec:conc}.

\section{Model and Methods}
\label{sec:model}

In order to constrain hadronic models for the observed extra spectral component
of GRB 090510, we simulate the photon emission
with photomeson-induced cascade processes
from ultrarelativistic outflows of bulk Lorentz factor $\Gamma$.
The numerical code is the same as that in \citet{asa09},
which was matured via a series of GRB studies \citep{asa05,asa06,asa07},
so that we omit the detailed explanation for the code.
As an optimistic case, which would lead to hadronic cascades,
we assume that protons are injected with a power law energy distribution
$\propto E_{\rm p}^{-2} \exp{(-E_{\rm p}/E_{\rm max})}$ above
10 GeV in the outflow rest frame at radii $R$ from the central engine.
The acceleration timescale of a proton is parameterized as
$t_{\rm acc}=\xi R_{\rm L}/c$, where $R_{\rm L}$ is the Larmor radius of the proton.
The maximum proton energy $E_{\rm max}$ is determined by
equating the acceleration timescale and the expansion timescale
of the outflow $t_{\rm exp}=R/(c\Gamma)$ or the cooling timescale due to
proton synchrotron or photomeson production.

We assume that the main low-energy component, which
is fitted by a Band function, is of leptonic origin \cite{abd09b},
which we do not discuss further.
The photon energy distribution in the outflow rest frame
is estimated from the time-averaged Band parameters from $T_0+0.5$ s
to $T_0+1.0$ s ($\varepsilon_{\rm peak}=3.9$ MeV, $\alpha=-0.58$,
$\beta=-2.83$, and the isotropic-equivalent luminosity
$L_\gamma=2.5 \times 10^{53}$ erg$/$s at $z=0.903$)
and the numerically obtained extra component originating from protons.
The energy distributions of photons and particles are simulated iteratively
until they converge to a self-consistent spectrum.
On a timescale comparable to $t_{\rm exp}$, the photon density and
the magnetic fields, etc., can be approximately taken as constant, and we
can neglect the emission from particles beyond this timescale after their
injection, because the photon density and the magnetic field decline.
The physical processes taken into account are 1) photon emission
processes of synchrotron and Klein-Nishina regime Compton scattering
for electrons/positrons, protons, pions, and muons,
2) synchrotron self-absorption for electrons/positrons,
3) $\gamma \gamma$ pair production, 4) photomeson production from protons and neutrons,
5) Bethe-Heitler pair production ($p + \gamma \to p + e^+ + e^-$), and
6) decays of pions, and muons.
The code can also output the spectra of neutrinos from pions and muons.

There are five model parameters: the proton acceleration parameter $\xi$,
the proton injection radius $R$, the bulk Lorentz factor $\Gamma$,
the energy density of the magnetic field $U_B$,
and the injection luminosity of accelerated protons $L_{\rm p}$.
The last two parameters are normalized by the photon energy density
of the Band component $U_\gamma=L_\gamma/(4 \pi c R^2 \Gamma^2)$ or
$L_\gamma$ itself as $U_B/U_\gamma$ and $L_{\rm p}/L_\gamma$.

\section{Hadronic Emission}
\label{sec:results}

Within the time interval between $T_0+0.5$ s and $T_0+1.0$ s,
two high-energy photons of 3.4 GeV and 31 GeV were detected.
These photons belong to the extra component of the prompt emission, whose spectrum
is well fitted by a power law of $-1.6$,
on which we concentrate here. (There is in addition a tail
extending to $\sim 200$ s, attributable to an afterglow, which we
do not discuss here.)
Since the photon statistics above GeV are not enough,
we do not attempt to carry out a detailed fitting of the spectrum.
Alternatively, we search for parameter sets that reproduce
a comparable fluence of the extra component to that of the Band component.

\begin{figure}[htb!]
\centering
\epsscale{1.0}
\plotone{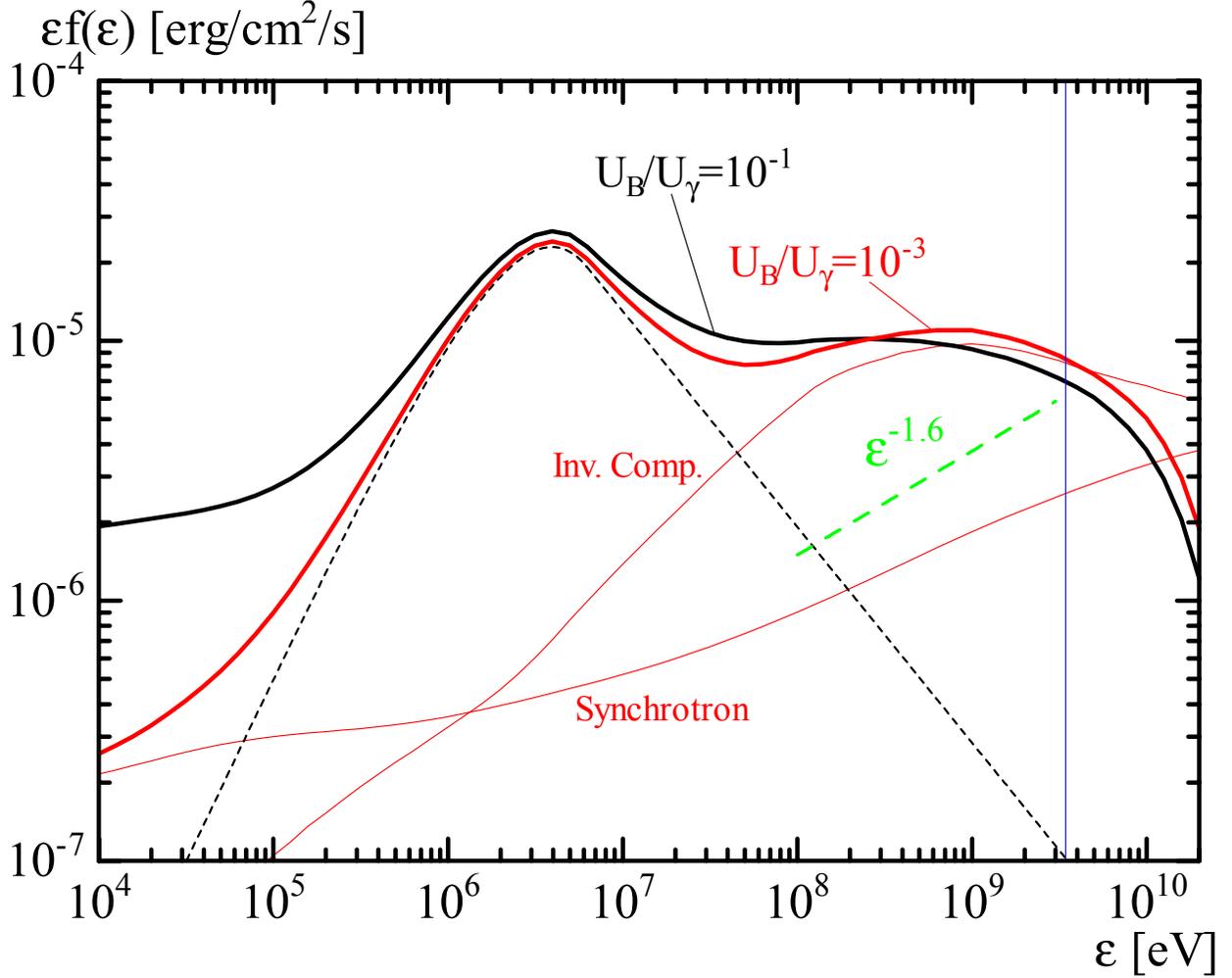}
\caption{
Photon spectra (bold curve) for $\xi=1$, $R=10^{14}$ cm, and $\Gamma=1500$.
The red curve is for $U_B/U_\gamma=10^{-3}$ and $L_{\rm p}/L_\gamma=200$,
and the black curve is for $U_B/U_\gamma=10^{-1}$  and $L_{\rm p}/L_\gamma=30$.
Fine red curves denote separately pair synchrotron and
inverse Compton without the absorption effects.
The fine dashed curve is the Band component.
The vertical blue line denotes the energy of the 3.4 GeV photon.
\label{fig:f1}}
\end{figure}

First, we consider photomeson cascades, for the most extreme case $\xi=1$.
Fig. \ref{fig:f1} shows two examples of photon spectra,
in which the flux of the hadronic component at 3.4 GeV
is comparable to the Band component.
The model spectrum with $U_B/U_\gamma=10^{-3}$ seems consistent with
the observed spectrum.
The second peak at $\sim$ GeV is mainly due to $\gamma \gamma$ pair absorption.
The assumed value of $L_{\rm p}/L_\gamma$ is 200, which is quite large,
so that the proton injection luminosity should be larger than
$5 \times 10^{55}$ erg$/$s in this case.
On the other hand, if we adopt a stronger magnetic field
such as $U_B/U_\gamma=10^{-1}$, the required amount of protons can be suppressed
to a lower value of $L_{\rm p}/L_\gamma=30$ ($L_{\rm p} \sim
7 \times 10^{54}$ erg$/$s).
However, the dominance of synchrotron radiation
from secondary pairs makes in this case the spectrum too soft below about 100 keV.
Such a large deviation from the Band function around 10 keV
is not seen in GRB 090510.

\begin{figure}[htb!]
\centering
\epsscale{1.0}
\plotone{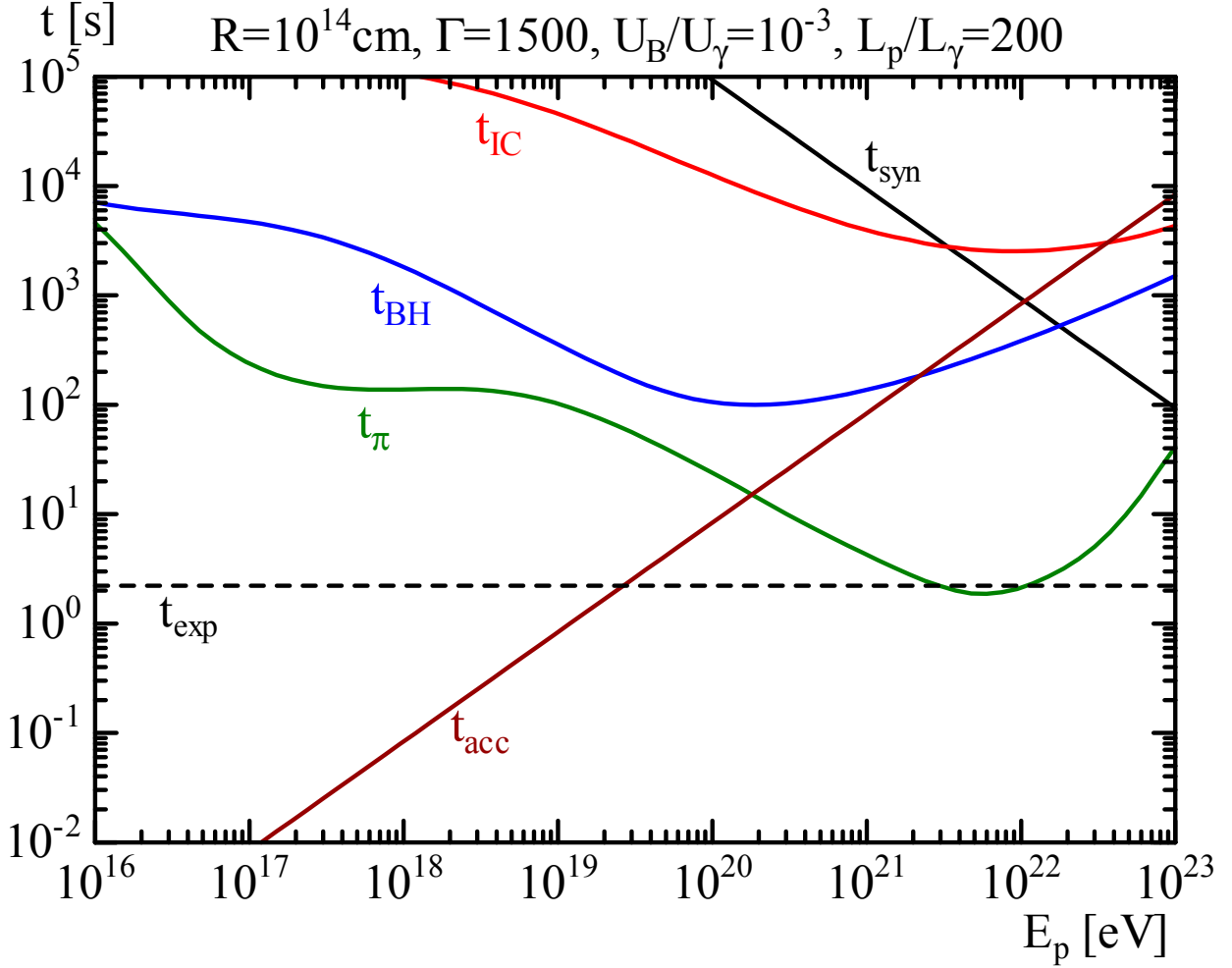}
\caption{
Various timescales for the model of $U_B/U_\gamma=10^{-3}$ in Fig. \ref{fig:f1}:
proton cooling timescales due to photomeson production ($t_\pi$),
Bethe-Heitler pair production ($t_{\rm BH}$), synchrotron emission ($t_{\rm syn}$),
and inverse Compton emission ($t_{\rm IC}$), proton acceleration timescale ($t_{\rm acc}$),
and expansion timescale of the outflow ($t_{\rm exp}$) in the outflow rest frame.
\label{fig:f2}}
\end{figure}

The various timescales for the case of $U_B/U_\gamma=10^{-3}$ are plotted
in Fig. \ref{fig:f2}.
Apparently, the maximum energy of protons
$E_{\rm max}$ is determined by the condition $t_{\rm exp}=t_{\rm acc}$.
The low efficiency of hadronic cascade is attributed to the much longer
cooling timescale due to photopion production $t_\pi$ than
the expansion timescale of the outflow $t_{\rm exp}$ even at $E_{\rm p}=E_{\rm max}$.
The target photons for pion production below $\sim 3 \times 10^{18}$ eV
are the Band component.
Above this energy, protons interact with UV-photons
below the observed energy range of Fermi, where secondary synchrotron photons dominate.
In our simulation the synchrotron self-absorption makes a spectral peak
of $\varepsilon N(\varepsilon)$ plot at $\sim 40$ eV, which corresponds
to the typical target photon energy for protons of $\sim 10^{22}$ eV,
where $t_\pi$ becomes minimum.
In spite of the enhancement of the pion production efficiency due to secondary photons,
we still need a large proton luminosity.
The curves in Fig. \ref{fig:f2} can be shifted by changes of $R$,
$\Gamma$, and $U_B/U_\gamma$.
The increase of $E_{\rm max}$ due to the decrease of $t_{\rm acc} \propto B^{-1}$
with increasing magnetic fields leads to a higher efficiency of photomeson production.
As shown in Fig. \ref{fig:f1}, however, we need to decrease the magnetic field
to values as low as $U_B/U_\gamma=10^{-3}$ in order to harden the spectrum,
since the Klein-Nishina effect is crucial.

Let us consider protons that interact with photons of $\varepsilon_{\rm peak}$.
The cooling timescale for such protons behaves as $t_\pi \propto n_\gamma^{-1}$,
where the photon density $n_\gamma=U_\gamma/(\varepsilon_{\rm peak}/\Gamma)$.
If we adopt a smaller value of $\Gamma$,
the pion production efficiency would be improved as
$t_{\rm exp}/t_\pi \propto R^{-1} \Gamma^{-2}$.
However, we should note that there is a lower limit to $\Gamma$, which is required to
make the source optically thin to GeV photons.
Given the photon luminosity and spectral shape,
this minimum Lorentz factor can be estimated as shown in the online supporting materials
in \citet{abd09,abd09b}.
The adopted value $\Gamma=1500$ in Fig. \ref{fig:f1}
is close to the lower limit for $R=10^{14}$ cm.
Even if we take a larger $R$, the lower limit of the Lorentz factor
$\propto R^{1/(\beta-3)}$ does not decrease drastically
(since $\beta \simeq -3$, $\Gamma_{\rm min} \propto R^{-1/6}$).
Although a smaller $R \ll 10^{14}$ cm and a slightly larger $\Gamma$
can reduce the required amount of protons,
the typical variability timescale
$R/(c \Gamma^2)$ in Fig. \ref{fig:f1} ($R=10^{14}$ cm and $\Gamma=1500$)
is already as small as $1.5$ ms.
The poor photon statistics in GRB 090510 cannot constrain well the variability
timescale, but the required proton energy may remain very large even for
$R/(c \Gamma^2) \ll 1$ ms
(note that in \citet{abd09b} the variability timescales
are estimated as $\sim 10$ ms, which
are adopted to obtain the minimum Lorentz factor).

\begin{figure}[htb!]
\centering
\epsscale{1.0}
\plotone{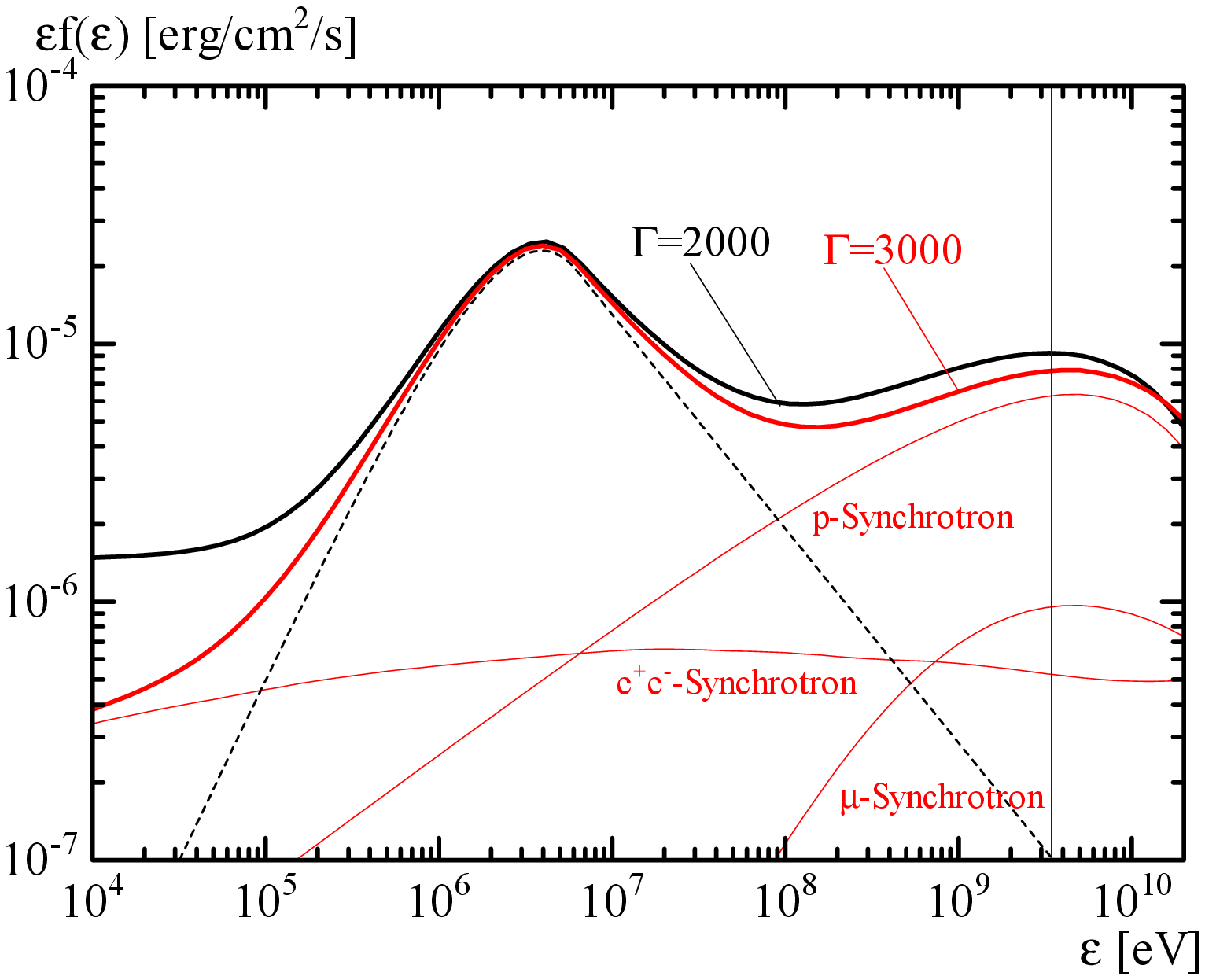}
\caption{
Photon spectra (bold curve) for $\xi=10^3$ and $R=10^{14}$ cm.
The red curve is for $\Gamma=3000$, $U_B/U_\gamma=300$, and $L_{\rm p}/L_\gamma=300$,
and the black curve is for $\Gamma=2000$, $U_B/U_\gamma=200$, and $L_{\rm p}/L_\gamma=100$.
Fine red curves denote separately proton synchrotron,
pair synchrotron, and muon synchrotron
without the absorption effects for $\Gamma=2000$.
The fine dashed curve is the Band component.
\label{fig:f3}}
\end{figure}

An alternative hadronic scenario is a proton synchrotron model
with very strong magnetic field.
In this model, we need to avoid the pair cascades initiated by photomeson production,
because the synchrotron spectrum from secondary pairs is too soft as discussed above.
Therefore, we adopt $\xi=10^3$ to suppress $E_{\rm max}$.
Examples for this model in Fig. \ref{fig:f3}
require the same order of magnitude ratios of $L_{\rm p}/L_\gamma$ as the pair cascade model.
Considering the typical energy of protons emitting GeV photons,
$\propto (B \Gamma)^{-1/2}$,
the efficiency of proton synchrotron can be enhanced by a decrease of $\Gamma$,
according to
$t_{\rm exp}/t_{\rm syn} \propto \Gamma^{-3} R^{-1/2} (U_B/U_\gamma)^{3/4}$.
However, even for $\Gamma=3000$ in Fig. \ref{fig:f3},
the pair cascades due to photomeson production
cannot be neglected in the low-energy regions.
The photon spectrum below 100 keV becomes too soft again for $\Gamma=2000$ owing to
the pair cascade.
If we can neglect the observed variability timescale $\lesssim 10$ ms,
it seems that a much larger $R$ and slightly smaller $\Gamma$ could decrease the required
luminosity of protons.
Under the conservative assumption of $U_B/U_{\rm p} \leq 1$, however,
a decrease of $U_{\rm p}$ leads to a low efficiency of proton synchrotron
due to the low $U_B$. Considering the dependence of $t_{\rm exp}/t_{\rm syn}
\propto \Gamma^{-3} R^{-1/2} (U_B/U_\gamma)^{3/4}$,
we cannot avoid the requirement of a large proton luminosity even in the
proton synchrotron scenario.

In the above, we have sought to reproduce the flux at 3.4 GeV
by photons originating from protons.
To extend the extra component as far as 31 GeV,
a much larger $\Gamma$ is required, which would decrease the efficiency 
of photopion production.
(for example, $L_{\rm p}/L_\gamma=1000$ is required to obtain
sufficient flux at 31 GeV for $\xi=1$, $\Gamma=3000$,
$U_B/U_\gamma=10^{-3}$, and $R=10^{14}$ cm).
Thus, we need an isotropic equivalent proton luminosity in excess of
$10^{55}$ erg$/$s in order to
explain the photon flux and hard spectrum of the extra component, independently
of the model details.

Another aspect of this burst is the reported $\sim 0.1$-0.2 s
delayed onset of the LAT high-energy photons relative to the GBM trigger.
This delayed onset is a common property of other {\it Fermi}/LAT GRBs
\citep[etc.]{abd09}.
Within the hadronic models, this delay could in principle be attributed to
the acceleration timescale of protons, which is limited
by the shell expansion timescale.
This would suggest that the timescale of the delayed onset
is comparable to the variability timescale, which may be comparable to
or shorter than 10 ms.
However, since the cooling timescale is much longer than
the shell expansion timescale in our models
(which is equivalent to the requirement $L_{\rm p}/L_\gamma \gg 1$),
we cannot deduce the delay timescale in this manner from the models considered here.
In order to explain the delayed onset,
we would need additional assumptions, an example of which
might be spatially different origins of the Band and the extra component.
However, with only two photons above 1 GeV, it is premature to engage here
in such additional modeling.

\begin{figure}[htb!]
\centering
\epsscale{1.0}
\plotone{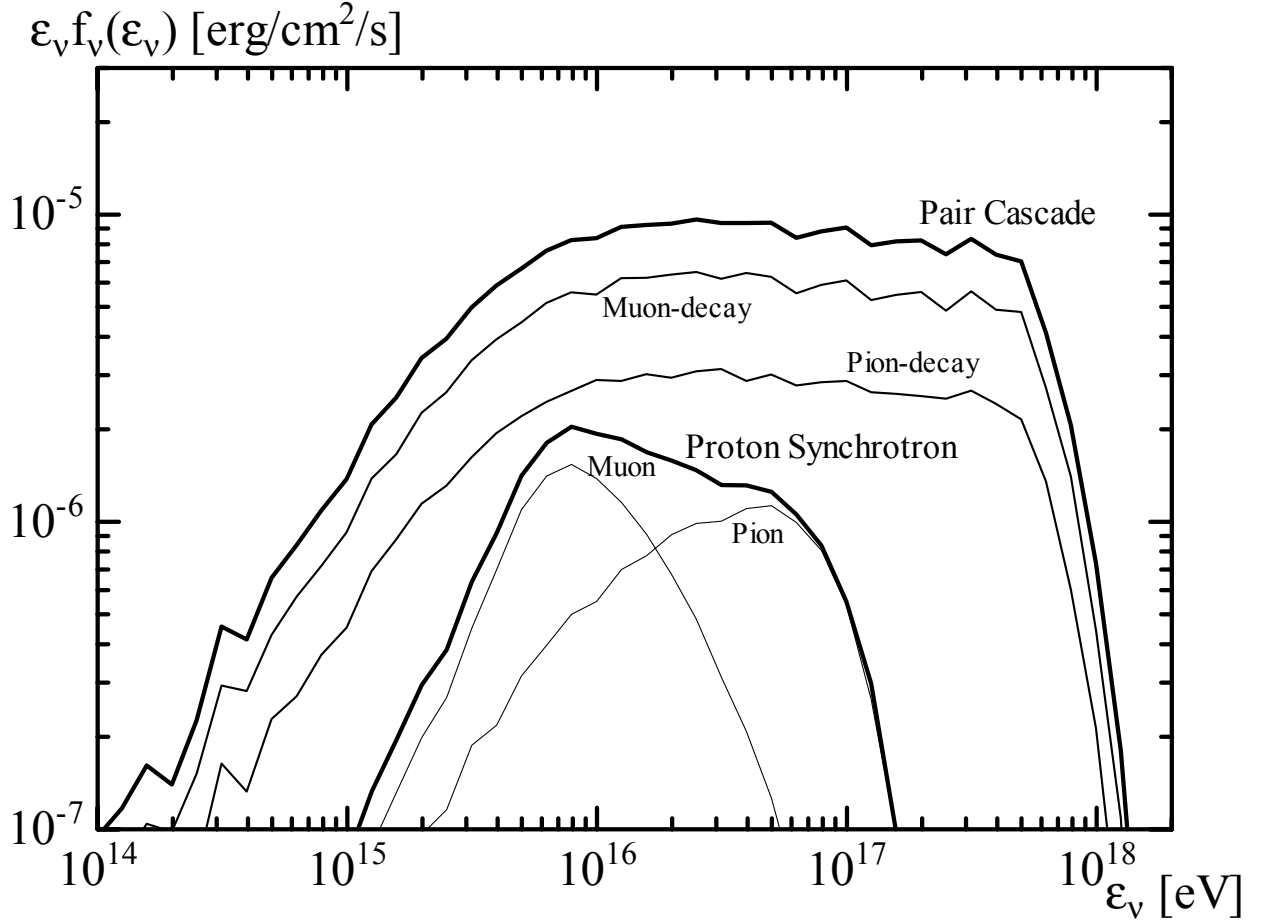}
\caption{
Neutrino spectra (black bold curves) for the pair cascade model
($\Gamma=1500$, $\xi=1$, $L_{\rm p}/L_\gamma=200$, $U_B/U_\gamma=10^{-3}$,
and $R=10^{14}$ cm) and the proton synchrotron model ($\Gamma=3000$,
$\xi=10^3$, $L_{\rm p}/L_\gamma=300$, $U_B/U_\gamma=300$,
and $R=10^{14}$ cm).
Fine curves denote separately neutrinos from pion decay
and muon decay.
\label{fig:f4}}
\end{figure}

In Fig. \ref{fig:f4}, we have also plotted the neutrino spectra (sum over all species)
obtained from the model for $U_B/U_\gamma=10^{-3}$ in Fig. \ref{fig:f1}
(photopion-induced pair cascade model)
and for $\Gamma=3000$ in Fig. \ref{fig:f3} (proton synchrotron model).
For the pair cascade case,
the very large proton luminosity leads to a neutrino luminosity comparable
to the photon luminosity.
The neutrino number flux is estimated as $\sim 6~{\rm km}^{-2}~{\rm s}^{-1}$
for each energy decade from $10^{14}$ to $10^{16}$ eV.
Considering the effective area of the IceCube detector \citep{abb09},
we may not expect neutrino detections from this burst
(the upper limit of the muon neutrino fluence for the ``naked-eye'' GRB 080319B
by the IceCube detector is $\sim 10^{-2} {\rm erg}~{\rm cm}^{-2}$
for $<2.2 \times 10^{15}$ eV).
For the proton synchrotron model, the strong magnetic field efficiently
cools pions and muons via synchrotron radiation so that the highest neutrino energies
are suppressed, compared to the pair cascade model.

\section{Conclusions}
\label{sec:conc}

The detection of GeV photons from the short GRB 090510 requires
a very large Lorentz factor, which leads to a low efficiency of
photomeson production.
Adopting the observed spectrum, we have simulated the photon emission
due to accelerated protons to produce the extra spectral component around GeV.
We have shown that this mechanism is capable of reproducing the flux
in this extra component, during the approximate time interval between the
trigger and $0.8$ s when this component is present.
However, because of the low efficiency of photopion production,
a very large proton isotropic-equivalent luminosity of $>10^{55}$ erg$/$s is required
to produce GeV photons from electron-positron pairs.
While we have assumed a power-law proton spectrum with an exponential cutoff
for simplicity, the energy range, in which photopion production predominantly contributes
to the secondary photon emissions, is narrow as shown in Fig. \ref{fig:f2}.
The required proton luminosity we suggest here
practically represents the required normalization of the proton flux
at this energy range ($\sim 10^{18}$ eV for the example in Fig. \ref{fig:f2}).
Even if we change the proton injection spectrum, the effect on the photon spectrum
would be negligible as long as the proton amount in that energy range is the same.
Fixing the proton amount in this energy range, of course,
steeper proton spectra enhance the total injection luminosity.
In principle, a lower proton luminosity associated with a stronger magnetic 
field could reproduce the flux around GeV, provided the spectral modifications
below 100 keV  due to secondary pairs are weak. This remains to be investigated,
but in the absence of some physical mechanism to suppress the emission of 
low-energy photons from secondary pairs, the spectrum predicted from such a 
strong field model would contradict the observations.
Also for the proton synchrotron model with $U_B/U_\gamma \gg 1$ discussed above,
the required proton isotropic-equivalent luminosity is $\gtrsim 10^{55}$ erg$/$s.

Such ``proton-dominated'' GRB models \citep[e.g.,][]{asa09} are favorable
for scenarios of ultra-high-energy cosmic-ray generation by GRBs,
but as in this case too, they require very high
isotropic-equivalent proton energies $\gtrsim 10^{55}$ erg/s, which are 
challenging for such hadronic models. However, we should note that even
the photon isotropic equivalent luminosities of quite a few bursts have been
found to be of that order of magnitude (e.g., GRB 080913, etc.), so that the
hadronic interpretation of the extra GeV component does not exceed such
observed photon luminosities in other bursts. Jet collimation effects can
significantly alleviate such energetic problems. Unfortunately, in the case of
GRB 090510 the afterglow is brief, and there is so far no evidence for or 
against any jet collimation.
Adopting $1/500 > 1/\Gamma$ as the collimation factor,
the required energy may strain a conventional NS-BH merger model,
but is not  huge.
The strain would be less on a core collapse
model, but this model has not solved the short-duration issue yet.
Therefore,
the extreme energetics of such hadronic outflow models remain a substantial challenge.

In order to distinguish models of internal shock origin including hadronic models
and afterglow origin \citep{ghi09} for GeV photons, better photon statistics
above GeV are desirable. Although the correlation function for this burst
shows a clear correlation between GBM (250 keV-3MeV) and LAT ($>0.1$ GeV) light curves
\citep{abd09b}, the typical photon number above GeV in a pulse is a few.
Closer GRBs than GRB 090510 at least by a factor of 3, from which we expect
more than 10 photons in a pulse, may strengthen the correlation,
but we should note that the GRB rate may decrease with increasing
photon statistics (the present detection rate with Fermi LAT is
$\sim 10$ GRBs $\mbox{yr}^{-1}$).

\begin{acknowledgments}
We thank the {\it Fermi} GBM/LAT collaborations, especially S. Razzaque,
C. Dermer, and J. Granot, for valuable discussions,
and acknowledge partial support from NSF PHY-0757155
and NASA NNX 08AL40G (P.M.).
Finally, we appreciate the anonymous referee.
\end{acknowledgments}


\end{document}